\documentclass[12pt,]{article}
\usepackage{lmodern}
\usepackage{amssymb,amsmath}
\usepackage{ifxetex,ifluatex}
\usepackage{fixltx2e} % provides \textsubscript
\ifnum 0\ifxetex 1\fi\ifluatex 1\fi=0 % if pdftex
  \usepackage[T1]{fontenc}
  \usepackage[utf8]{inputenc}
\else % if luatex or xelatex
  \ifxetex
    \usepackage{mathspec}
  \else
    \usepackage{fontspec}
  \fi
  \defaultfontfeatures{Ligatures=TeX,Scale=MatchLowercase}
\fi
\IfFileExists{upquote.sty}{\usepackage{upquote}}{}
\IfFileExists{microtype.sty}{%
\usepackage{microtype}
\UseMicrotypeSet[protrusion]{basicmath} % disable protrusion for tt fonts
}{}
\usepackage[margin=1in]{geometry}
\usepackage{hyperref}
\hypersetup{unicode=true,
            pdftitle={Spatiotemporal modelling of sea duck abundance: implications for marine spatial planning},
            pdfborder={0 0 0},
            breaklinks=true}
\usepackage{graphicx,grffile}
\makeatletter
\def\maxwidth{\ifdim\Gin@nat@width>\linewidth\linewidth\else\Gin@nat@width\fi}
\def\maxheight{\ifdim\Gin@nat@height>\textheight\textheight\else\Gin@nat@height\fi}
\makeatother
\setkeys{Gin}{width=\maxwidth,height=\maxheight,keepaspectratio}
\IfFileExists{parskip.sty}{%
\usepackage{parskip}
}{% else
\setlength{\parindent}{0pt}
\setlength{\parskip}{6pt plus 2pt minus 1pt}
}
\ifx\paragraph\undefined\else
\let\oldparagraph\paragraph
\renewcommand{\paragraph}[1]{\oldparagraph{#1}\mbox{}}
\fi
\ifx\subparagraph\undefined\else
\let\oldsubparagraph\subparagraph
\renewcommand{\subparagraph}[1]{\oldsubparagraph{#1}\mbox{}}
\fi

\let\rmarkdownfootnote\footnote%
\def\footnote{\protect\rmarkdownfootnote}

\usepackage{titling}

  \title{Spatiotemporal modelling of sea duck abundance: implications for marine
spatial planning}
  \pretitle{\vspace{\droptitle}\centering\huge}
  \posttitle{\par}
  \author{}
  \preauthor{}\postauthor{}
  \date{}
  \predate{}\postdate{}

\usepackage[utf8]{inputenc} 
\usepackage{hanging}
\usepackage{pdfpages}
\usepackage[document]{ragged2e}
\usepackage{float}
\usepackage{setspace}

\begin{document}
\maketitle

\doublespacing

Adam D. Smith\textsuperscript{1,}\footnotemark[6], Benjamin
Hofner\textsuperscript{2}, Jason E. Osenkowski\textsuperscript{3}, Taber
Allison\textsuperscript{4}, Giancarlo Sadoti\textsuperscript{5}, Scott
R. McWilliams\textsuperscript{1}, and Peter W. C.
Paton\textsuperscript{1}

\textsuperscript{1}Department of Natural Resources Science, University
of Rhode Island, Kingston, Rhode Island 02881 USA

\textsuperscript{2}Department of Medical Informatics, Biometry and
Epidemiology, Friedrich-Alexander-University Erlangen-Nuremberg,
Waldstraße 6, 91054 Erlangen, Germany

\textsuperscript{3}Rhode Island Department of Environmental Management,
77 Great Neck Road, West Kingston, Rhode Island 02892 USA

\textsuperscript{4}American Wind Wildlife Institute, 1110 Vermont
Avenue, NW, Suite 950, Washington, D.C. 20005 USA

\textsuperscript{5}Department of Geography, University of Nevada, Reno,
Nevada, 89557 USA

\bigskip

RUNNING HEAD: Model-based distribution and abundance

\bigskip

\footnotetext[6]{Present address: United States Fish and Wildlife Service, National Wildlife Refuge System, Inventory and Monitoring Branch, 135 Phoenix Rd., Athens, GA 30605 USA; avianmigration@gmail.com}

\newpage

\doublespacing

\textbf{Summary}

\begin{enumerate}
\def\labelenumi{\arabic{enumi}.}
\item
  Effective marine spatial plans require information on the distribution
  and abundance of biological resources that are potentially vulnerable
  to anthropogenic change. In North America, spatially-explicit
  abundance estimates of marine birds are necessary to assess potential
  impacts from planned offshore wind energy developments (OWED). Sea
  ducks are particularly relevant in this context as populations of most
  North American species are below historic levels and European studies
  suggest OWEDs.
\item
  We modelled species occupancy using a generalized additive model and
  conditional abundance with generalized additive models for location,
  scale, and shape; the models were subsequently combined to estimate
  unconditional abundance. We demonstrate this flexible, model-based
  approach using sea ducks (Common Eider {[}\emph{Somateria
  mollissima}{]}, Black Scoter {[}\emph{Melanitta americana}{]}, Surf
  Scoter {[}\emph{M. perspicillata}{]}, White-winged Scoter {[}\emph{M.
  deglandi}{]}, and Long-tailed Duck {[}\emph{Clangula hyemalis}{]}) in
  Nantucket Sound, Massachusetts, USA, which supports some of the
  largest concentrations of wintering sea ducks in eastern North America
  and where a 454-MW OWED is proposed.
\item
  Spatiotemporal effects were the dominant explanatory features in sea
  duck occupancy and conditional abundance models. Biophysical
  covariates also influenced occupancy (i.e., highest at intermediate
  sea surface temperatures and in areas with coarser sea floor
  sediments), but their effects on conditional abundance were less
  consistent among species. Spatially-explicit abundance estimates
  suggested that while the planned OWED in Nantucket Sound may displace
  some sea ducks from potential foraging habitat, most occurred in areas
  away from the proposed OWED.\\
\item
  \emph{Synthesis and application}. Our approach to species distribution
  and abundance modelling offers several useful features including (1)
  the ability to model all conditional distribution parameters as a
  function of covariates, (2) integrated variable reduction and
  selection among many covariates, (3) integrated model selection, and
  (4) the efficient incorporation of smooth effects to capture
  spatiotemporal trends poorly explained by other covariates. This
  modelling approach should prove useful for marine spatial planners in
  siting OWEDs while considering key habitats and areas potentially
  vulnerable to anthropogenic stressors. Moreover, the approach is
  equally suitable for terrestrial or aquatic systems.
\end{enumerate}

\bigskip


\newpage

\subsection{Introduction}\label{introduction}

Marine spatial plans (MSP) can provide a rigorous framework to protect
marine ecosystems from a variety of anthropogenic stressors (Polasky
\emph{et al.} 2008; Foley \emph{et al.} 2010). A key element of MSPs is
accurate spatially-explicit estimates of the abundance and distribution
of biological resources (Douvere 2008; Punt \emph{et al.} 2009; Bradbury
\emph{et al.} 2014). Along the northwest Atlantic Coast, numerous
offshore wind energy developments (OWED; Breton \& Moe 2009; Musial \&
Ram 2010) are proposed, including a 30-MW facility in Rhode Island Sound
to be the first U.S. operational OWED by fall 2016. Therefore there is a
pressing need to develop spatially-explicit models of key biological
resources that can assist marine spatial planners in siting OWEDs while
considering marine ecosystem integrity.

Understanding the distribution and abundance of marine birds represents
a particular challenge for MSPs because the environmental factors
associated with their distribution and abundance remain notably
understudied (Zipkin \emph{et al.} 2010; Bowman \emph{et al.} 2015;
Flanders \emph{et al.} 2015). Sea ducks (Tribe: Mergini) are
particularly important to consider in the context of marine spatial
planning because populations of most North American species are below
historic levels and there is considerable uncertainty in population
trajectories (Bowman \emph{et al.} 2015). Moreover, evidence from Europe
suggests that OWEDs may negatively impact sea duck populations (Larsen
\& Guillemette 2007; Langston 2013; Bradbury \emph{et al.} 2014).

Model-based approaches are a useful framework to generate
spatially-explicit estimates of animal abundance and changes in animal
distribution independent of abundance (Borchers \emph{et al.} 2002;
Certain \& Bretagnolle 2008; Nur \emph{et al.} 2011; Kinlan \emph{et
al.} 2012; Winiarski \emph{et al.} 2014). However, modelling the spatial
ecology of marine organisms such as sea ducks presents a variety of
analytical challenges. First, the spatiotemporal distribution of marine
organisms can be highly irregular and surveys often produce
zero-inflated (Martin \emph{et al.} 2005; Cunningham \& Lindenmayer
2005), overdispersed (Richards 2008) count data that may vary with
biophysical features in complex, non-linear ways (Austin 2007). Second,
an adequate modelling approach usually must identify a small suite of
important covariates among many potentially correlated covariates while
avoiding overfitting, a process that often is complicated by data from
repeated surveys that regularly exhibit spatial and temporal
autocorrelation (Hoeting 2009).

We demonstrate a flexible model-based approach to predict animal
distribution and abundance using aerial transect surveys of sea ducks in
Nantucket Sound, Massachusetts, USA. We implemented generalized additive
models (GAMs; Hastie \& Tibshirani 1990; Wood 2006) and their recent
extension (GAMLSS - GAMs for location, scale, and shape; Rigby \&
Stasinopoulos 2005) in a gradient descent boosting framework (Friedman
2001). GAMLSS methods extend GAMs to allow all parameters of the
conditional response distribution to be modelled as a function of
relevant covariates (Rigby \& Stasinopoulos 2005). This approach
provides spatially-explicit insights into the covariates associated with
sea duck abundance and its variability. The boosting algorithm
accommodates the inclusion of various effect types (e.g., linear,
smooth, spatial, spatiotemporal, and random effects; Hofner \emph{et
al.} 2014) and correlated covariates (i.e., parameter estimates are
regularized), identifies the most relevant subset among potentially many
covariates (i.e., variable selection), and evaluates competing
representations of continuous covariates (e.g., linear vs.~nonlinear
effects; Kneib \emph{et al.} 2009; Maloney \emph{et al.} 2012). Our
specific objectives were to demonstrate a flexible, model-based approach
that (1) provides spatially- and temporally-explicit estimates of sea
duck abundance and distribution while accommodating many of the
challenges inherent in animal survey data; and (2) describes the
associations between biophysical features and sea duck distribution and
abundance in one of the most important migrating and wintering areas for
sea ducks in the western Atlantic.

\subsection{Materials and methods}\label{materials-and-methods}

\emph{Study area and species.}---We conducted fieldwork throughout
Nantucket Sound in Massachusetts, USA (Figure 1). Our study area
encompassed ca. 1,500 km\textsuperscript{2}, was relatively shallow
(generally \textless{} 20 m deep), and included some of the most
important sea duck wintering habitat in the western Atlantic (Silverman
\emph{et al.} 2013). The primary species of sea ducks found in Nantucket
Sound were Common Eider (\emph{Somateria mollissima}; hereafter eider),
Black Scoter (\emph{M. americana}), Surf Scoter (\emph{M.
perspicillata}), and White-winged Scoter (\emph{Melanitta deglandi}),
and Long-tailed Duck (\emph{Clangula hyemalis}). Additionally,
approximately 62 km\textsuperscript{2} of of Horseshoe Shoal in
northwestern Nantucket Sound is fully permitted for OWED (Figure 1;
Santora \emph{et al.} 2004), which may affect sea duck use of this
important wintering area (Flanders \emph{et al.} 2015).

\emph{Aerial strip transect surveys.}---During the winters (late October
to mid-April) of 2003-2005, we conducted 30 (2003/2004: 13, 2004/2005:
10, 2005/2006: 7) standardized aerial strip-transect surveys (Flanders
\emph{et al.} 2015) throughout an 1,100 km\textsuperscript{2} study area
in Nantucket Sound (Figure 1). Surveys occurred primarily from November
- March (\emph{n} = 27), with occasional October (\emph{n} = 1) or April
(\emph{n} = 2) surveys. During each survey, we flew along 15 parallel
north-south transects, separated by ca. 2.5 km (Figure 1), using a
high-wing, twin-engine aircraft at an altitude of 152 m and speed of 167
km/h (90 kts). This altitude allowed us to identify most birds on the
sea surface and reduced disturbance (i.e., flushing birds to another
part of the study area and, thus, potential double counting). Surveys
occurred only on days with wind speeds \(\leq\) 15 kts and good
visibility (\textgreater{}15 km). Surveys, typically 2.5 h in duration,
occurred between 0900 to 1600 to avoid post-dawn and pre-sunset
movements by ducks (e.g., Davis 1997) and to reduce glare due to low sun
angles.

Observers monitored the sea surface within a ca. 90-m wide transect on
both sides of the plane; we thus sampled approximately 6\% (median; 68.4
km\textsuperscript{2}) of the study area during a survey. Observers
geographically referenced the number and species of ducks using the
plane's onboard GPS. We subsequently consolidated counts for each
species (eider and Long-tailed Duck) or species group (scoter) into
2.25km\textsuperscript{2} segments (Figure 1); this resolution (1.5 km x
1.5 km) corresponded approximately to the coarsest level of resolution
of biophysical covariates (see below).

\emph{Modelling approach.}---We related spatiotemporal variation in sea
duck occupancy and abundance to potentially relevant biophysical and
spatiotemporal covariates. We applied a negative binomial hurdle model
that first modelled (1) the probability of occurrence of at least one
individual (hereafter, the occupancy model) observed on the combined 180
m wide transect within a given segment using a logistic regression model
and then (2) the abundance of sea ducks observed on the transect in that
segment conditional on their presence (hereafter, the count model) using
a truncated negative binomial model. The prevalence of zero counts
(e.g., 75\% of eider segment observations) prompted the use of a hurdle
model, which we fit separately for each sea duck species group (Zipkin
\emph{et al.} 2010). GAMs and GAMLSS accommodated potential nonlinear
effects of covariates on sea duck occupancy and the conditional mean and
overdispersion of sea duck abundance, respectively.

We implemented GAMs and GAMLSS in a component-wise functional gradient
descent boosting framework. We first computed the negative gradient of a
user-specified loss function, which can be viewed as working residuals.
We used as the loss functions the (negative) binomial log-likelihood for
occupancy models and the (negative) truncated negative binomial
log-likelihood for count models.

Next step, we fit user-specified functional forms of the covariates
relative to the response, called base-learners (Hofner \emph{et al.}
2014), separately to the negative gradient and a fraction of the single
best fitting base-learner was added to the current model fit. The
negative gradient was then reevaluated at the current model fit and the
procedure iterated for a user-specified number of iterations,
\emph{\(m_\text{stop}\)} (see Bühlmann \& Hothorn 2007 for details). We
fit boosted GAMLSS models similarly, although in each iteration the
negative gradient was computed separately with respect to each GAMLSS
parameter (i.e., mean and overdispersion) while holding the other
parameter constant. We then fit base-learners to the resulting
parameter-specific negative gradient vector and model updates were
computed separately for each GAMLSS parameter as for boosted GAMs (see
Mayr \emph{et al.} 2012 for details).

As only the single best-fitting base-learner was selected in each
iteration, the algorithm integrated intrinsic selection of the most
relevant covariates and their functional form (i.e., some base-learners
were never selected). Variable selection was further fostered by
stopping the algorithm prior to convergence to maximum likelihood
estimates to maximize predictive accuracy while avoiding model
overfitting (early stopping; Mayr \emph{et al.} 2012; Maloney \emph{et
al.} 2012). We used 25-fold subsampling to determine the optimal
stopping iteration for each model by randomly drawing (without
replacement) 25 samples of size n/2 from the original data. We used the
selected sample to estimate the model and the balance of the data in
each sample to determine the out-of-bag prediction accuracy (empirical
risk) measured by the negative log-likelihood of each model; the optimal
stopping iteration (\(\hat{m}_\text{stop}\)) was the iteration with the
lowest average empirical risk. In boosted GAMLSS models, we used
multi-dimensional subsampling to determine the stopping iteration for
each GAMLSS parameter while allowing for potentially different model
complexities in the parameters (see Hofner \emph{et al.} 2015b).

Despite these agreeable features, boosting methods typically produce
``rich'' models relying to some extent on many base-learners (Hofner
\emph{et al.} 2015a). Thus, we applied stability selection (Meinshausen
\& Bühlmann 2010; Shah \& Samworth 2013) to identify the most commonly
selected base-learners, and thus covariates, in each model while
preserving an upper bound of \(\alpha\) \(\approx\) 0.06 for the
per-comparison error rate (see Appendix S1 and Hofner \emph{et al.}
2015a for details).

\emph{Covariates.}---We evaluated biophysical covariates expected to
influence the distribution and abundance of sea ducks and their benthic
prey (see Appendix S2), as we were unaware of data related directly to
the distribution of preferred prey (e.g., mollusks and crustaceans).
Additionally, we included interactions that allowed the effects of two
covariates to vary over time within a given winter (see Appendix S2). We
standardized (i.e., mean centered and scaled) all continuous covariates
and developed a custom function to visualize the spatial and temporal
distribution of covariates in Nantucket Sound (Appendix S3).

For each continuous covariate, we specified two base-learners: a linear
base-learner and a base-learner for the smooth deviation from the linear
effect via penalized splines (i.e., P-splines; Eilers \& Marx 1996;
Fahrmeir \emph{et al.} 2004; Schmid \& Hothorn 2008). This allowed the
algorithm to select between no effect, a linear effect, or a smooth
effect for each covariate. For categorical covariates, we included
\emph{K} - 1 (dummy-coded) linear base-learners for a \emph{K}-level
covariate. To address potential spatial autocorrelation, we included a
smooth surface function of the spatial coordinates of segment centers
(Kneib \emph{et al.} 2008); this surface comprised linear base-learners
for the easting and northing, their linear interaction, and a penalized
nonlinear tensor product P-spline (Kneib \emph{et al.} 2008, 2009;
Maloney \emph{et al.} 2012). We also allowed this surface to vary over
time within a winter.

The decomposition of continuous covariates into centered linear and
penalized orthogonal nonlinear base-learners allowed us to weight
base-learners equally and thus allowed unbiased model choice (i.e.,
prevented the preferential selection of smooth base-learners; Kneib
\emph{et al.} 2009; Hofner \emph{et al.} 2011). We restricted each
base-learner to a single degree of freedom and omitted the intercept
term from each base-learner.

The occupancy and count models had the following structure (see Appendix
S2 for descriptions of abbreviated covariates):

\begin{equation}
\begin{aligned}
g(\cdot) =  & int + time + f(time) + SSTw + f(SSTw) + SSTm + f(SSTm) + SSTrel +\\
     & f(SSTrel) + SSTrel \cdot time + f(SSTrel, time) + SBT + f(SBT) + NAOw + \\
     & depth + f(depth) + depth \cdot time + f(depth, time) + d2land + f(d2land) +\\
     & chla + f(chla) + cdom + f(cdom) + f(cdom, chla) + meanphi + f(meanphi) +\\
     & SAR + f(SAR) + tidebmean + f(tidebmean) + tidesd + f(tidesd) + strat +\\
     & f(strat) + ferry + y2004 + y2005 + xkm + ykm + xkm \cdot ykm + \\
     & f(xkm, ykm) + xkm \cdot time + ykm \cdot time + xkm \cdot ykm \cdot time + \\
     & f(xkm, ykm) \cdot time + obs\_window + f(obs\_window).  
\end{aligned}
\end{equation}

In our occupancy models, \(g(\cdot)\) is \(g(\pi_{\text{sea ducks}})\),
the occupancy probability of a given duck species and \(g\) is the logit
link. In our count models, \(g(\cdot)\) took two forms within the GAMLSS
framework -- the (conditional) mean count of sea ducks,
\(g(\mu_{\text{sea ducks}})\), and the (conditional) overdispersion in
sea duck counts, \(g(\sigma_{\text{sea ducks}})\); \(g\) is the log link
in both cases. Base-learners denoted as \(f(\cdot)\) indicate the
penalized nonlinear deviations from the corresponding linear
base-learner. The explicit intercept (\emph{int}) was a necessary
byproduct of our decomposition of base-learners (see Kneib \emph{et al.}
2009; Hofner \emph{et al.} 2011). We included \emph{obs\_window}, our
measure of survey effort (see Appendix S2), as a covariate because small
values in some segments impaired model estimability when treating
\emph{obs\_window} as an offset.

Subsequent to their independent fitting, we consolidated occupancy and
conditional count models (see Equation 6 in Zeileis \emph{et al.} 2008)
to generate spatially-explicit estimates of unconditional sea duck
abundance. These estimates were used to evaluate the approximate
explanatory power of our final models using a pseudo
\emph{R}\textsuperscript{2} measure of the explained variation
(Nagelkerke 1991; Maloney \emph{et al.} 2012).

All analyses were conducted in R (Version 3.1.3; R Core Team 2014) with
the add-on packages gamboostLSS (Hofner \emph{et al.} 2015c,b), mboost
(Hothorn \emph{et al.} 2010, 2015), and stabs (Hofner \& Hothorn 2015).
The data and code for reproducing this manuscript and analyses are given
as an online electronic supplement at
\url{http://github.com/adamdsmith/NanSound_JAppEcol}.

\subsection{Results}\label{results}

We modelled occupancy and conditional count independently for each
species. Bootstrapped empirical risk suggested that occupancy models for
all species converged to the maximum likelihood estimates (see Appendix
S4). Conversely, bootstrapped empirical risks prescribed early stopping
for both the conditional mean and overdispersion parameter in all count
models (see Appendix S4). Final occupancy models and conditional count
models included only a subset (12\% to 38\%) of the 48 base-learners
initially specified for selection (see equation 1). Occupancy models
generally contained more covariates and their interactions (8-10 of 23)
than did count models (3-6 of 23), particularly among stably selected
covariates and their interactions (Figure 2, see also Appendices S5 -
S7).

\emph{Sea duck occupancy.}---The covariates associated with occupancy
were relatively consistent among species (Figure 2). The influence of
univariate effects on the response is reflected in the range of the
effect over the Y-axis and, due to standardization, can be compared
among species and covariates within a model. For example, monthly sea
surface temperature (\emph{SSTm}) associated more strongly with eider
occupancy than did distance to land (\emph{d2land}) because it spanned a
larger range of the Y-axis (Figure 2). In contrast, monthly sea surface
temperature (\emph{SSTm}) associated much more strongly with occupancy
of Long-tailed Duck than eider and scoter for the same reason (Figure
2). Covariate interactions, illustrated with bivariate plots, are
similarly comparable within a model. Only the general association (i.e.,
positive or negative) with the additive predictor is given for factor
variables. Comparing univariate, bivariate, and categorical effects is
accomplished using the detailed covariate plots for eider, scoters, and
Long-tailed Duck (Appendices S5 - S7, respectively).

Spatiotemporal effects (i.e., the \emph{xkm}-\emph{ykm} location of
segments and the change over time within winter {[}\emph{time}{]}) were
the dominant explanatory features in occupancy models , although these
patterns varied considerably among species (Figure 2; see Day of season,
Northing x Easting). Occupancy increased, but at a decreasing rate, with
survey effort (\emph{obs\_window}) in a given segment, as well as at
intermediate monthly sea surface temperature (\emph{SSTm}), greater
distances from land (\emph{d2land}), and in areas with coarser sediments
(i.e., smaller \emph{meanphi}; Figure 2). Eider occupancy associated
negatively with chromomorphic dissolved organic material (\emph{cdom})
and positively with sea floor surface area relative to planimetric area
(\emph{SAR}; our measure of the topographic variability of the sea
floor; Figure 2), whereas scoter occupancy likewise related to
\emph{SAR} and \emph{cdom}, but in the opposite direction in both cases
(Figure 2). Scoter occupancy was modestly greater in deeper waters
(\emph{depth}), whereas Long-tailed Duck occupancy was greatest in
shallow waters during early winter and in deeper waters during later
winter (Figure 2; \emph{depth} x \emph{time} covariate). Other effects
were relatively minor and inconsistent among species.

The strong spatial effects (\emph{xkm}-\emph{ykm}) on occupancy resulted
in distinct spatial patterns of occupancy among species (Figure 3, top
row) despite the relative similarity of occupancy associations with
biophysical covariates. Occupancy was typically highest for eider in
northwest and southwest Nantucket Sound, in interior Nantucket Sound for
scoter, and in northeast and south Nantucket Sound for Long-tailed Duck
(Figure 3, top row). All species tended to avoid the western edge of the
Sound northeast of Martha's Vineyard. Generally, the areas of highest
occupancy exhibited the lowest relative variability (Figure 3, bottom
row), defined as the median absolute deviation (MAD) of occupancy
relative to median occupancy within a segment (a measure analogous to
the coefficient of variation).

\emph{Sea duck conditional abundance and overdispersion.}---Spatial
effects (\emph{xkm}-\emph{ykm}) were the dominant explanatory feature of
conditional abundance estimates for scoters and Long-tailed Duck, but
not for eider (Figure 2). In contrast to their occupancy model, scoter
conditional abundance decreased with increasing sediment grain size
(\emph{meanphi}). Additionally, the relationships between eider
conditional abundance and dissolved organic material (\emph{cdom}) and
sea floor topography (\emph{SAR}; Figure 2) were more complex than with
eider occupancy. The conditional abundance of eider and scoter was also
associated with relatively warm or cool sea surface temperatures
(\emph{SST\textsubscript{rel}}; Figure 2). Biophysical covariates
associated with Long-tailed Duck conditional abundance exhibited general
agreement with their corresponding occupancy model.

Spatially-explicit patterns of median conditional abundance (Figure 4,
top row) did not necessarily reflect occupancy patterns (cf.~Figure 3,
top row). Some areas of Nantucket Sound exhibited mutually high
conditional abundance and occupancy for a given species (e.g., eider in
the southwest, scoter in the interior, and Long-tailed Duck in parts of
the northeast). However, conditional abundance was low despite
relatively high occupancy in some areas (e.g., eider in the northeast
and Horseshoe Shoal, scoter in the northeast and southeast, and
Long-tailed Duck along the northern margin). Conversely, other areas of
Nantucket Sound exhibited lower occupancy but sea ducks, when present,
were more abundant (e.g., eider along the eastern margin, and scoter and
Long-tailed Duck in the southwest). As in occupancy models, sea ducks
were relatively absent from the middle-western margin of Nantucket Sound
(i.e., northeast of Martha's Vineyard; see Figure 1). In contrast to sea
duck occupancy, however, areas of highest conditional sea duck abundance
typically exhibited the highest relative variability over time (Figure
4, bottom row).

Overdispersion in conditional sea duck abundance also varied with
biophysical covariates, although there was less consistency in the
associated covariates among species (Figure 2; see also Appendices S5 -
S7). Variability (i.e., overdispersion) in sea duck counts was
heterogeneous in space (Appendix S8; Figure S8.1, top row) and time
(Appendix S8; Figure S8.1, bottom row), particularly for eider and
scoter (as indicated by the magnitude of the overdispersion parameter
values).

\emph{Expected sea duck abundance.}---Consolidated occupancy and
conditional count models provided estimates of unconditional sea duck
abundance in the study area over the survey period (Figure 5, top row).
Final models of expected sea duck abundance explained moderate amounts
of variation in observed counts of eider, scoter, and Long-tailed Duck
(pseudo \emph{R}\textsuperscript{2} = 0.31, 0.48, and 0.32,
respectively). Sea duck species exhibited relatively distinct patterns
of abundance in Nantucket Sound. Eider were consistently most abundant
in southwestern Nantucket Sound and relatively abundant in the
northeastern part of the study area but less consistently so as
evidenced by the relatively high MAD/median abundance over time (Figure
5). Scoter were also most abundant, with occasional extremely large
flocks, in southwestern Nantucket Sound, although this also represented
the area of highest relative variation in scoter abundance; relatively
high abundances of scoter also occurred in central Nantucket Sound
(Figure 5). Long-tailed Ducks were consistently most abundant in
northeastern Nantucket Sound, as well as along its southern margin
(Figure 5). No species' highest abundances occurred in the OWED
permitted Nantucket Shoal area, although expected eider and scoter
abundances were consistently elevated in some parts of the Shoal (west
and southeast, respectively; Figure 5).

We compared the total count of each sea duck species observed in aerial
strip transects with the corresponding estimated total abundance in
surveyed segments for each of the 30 aerial surveys (Figure 6). Our
models tended to overestimate sea duck abundance when actual numbers
were relatively low, although overestimation was typically less than an
order of magnitude. Additionally, scoter abundance was occasionally
extreme relative to typical counts and somewhat prone to underestimation
during these extreme counts. Nonetheless, the general adherence of
observed and predicted abundance to a line of unit slope indicated that
it may be reasonable to estimate sea duck abundance for the entire study
area based on observed sea duck densities in transects (Figure 6).

\emph{Temporal dynamics in wintering sea ducks.}---The MAD/median
estimates (bottom rows of Figures 3-5, Figure S8.1 in Appendix S8) show
that our spatially-explicit estimates of occupancy, abundance, and
overdispersion invariably change over time, either explicitly via the
selection of a within- or among-winter temporal effect (\emph{time} and
\emph{y2004}/\emph{y2005}, respectively) or implicitly via the selection
of biophysical covariates that change within or among winters. The
temporal dynamics of the wintering sea duck system in Nantucket Sound
was one of its most striking attributes (see an animation for scoter
occupancy and abundance in Appendix S9).

\subsection{Discussion}\label{discussion}

\emph{Utility of a boosted GAMLSS modelling framework.}---We
demonstrated a flexible model-based approach to evaluate the
environmental associations of species distribution and abundance based
on multiyear replicated surveys. This approach proved particularly
useful for sea ducks that exhibited considerable within and between year
variation in their spatial distribution and abundance (Zipkin \emph{et
al.} 2010; Winiarski \emph{et al.} 2014; Flanders \emph{et al.} 2015).
The boosted GAMLSS framework offered several useful features including
(1) the ability to model all conditional distribution parameters (e.g.,
conditional mean and overdispersion) as a function of covariates, (2)
integrated variable reduction and selection among many covariates, and
(3) integrated model selection via the simultaneous consideration of
competing functional covariate forms (e.g., linear vs.~non-linear).
Additionally, this framework allowed us to incorporate smooth effects to
efficiently account for spatiotemporal trends in the data that were
poorly explained by other covariates and to identify those covariates
and their functional forms most consistently associated with animal
distribution and abundance (via stability selection). Moreover, the
approach is equally suitable for terrestrial or aquatic systems.

\emph{The importance of spatial scale.}---The distribution and abundance
of species is often spatially and temporally dynamic because it is often
driven by biophysical covariates that may furthermore differ in
importance depending on spatial scale (Johnson 1980; Johnson \emph{et
al.} 2004, 2006). Considering our sea duck example, a larger-scale
occupancy model developed by Flanders \emph{et al.} (2015) suggested
that eiders were relatively uniformly distributed across Nantucket
Sound, whereas our higher resolution abundance models found that eiders
were concentrated in southwestern and central, eastern areas within
Nantucket Sound. Indeed, the spatiotemporal patterns of unconditional
sea duck abundance in Nantucket Sound largely reflect patterns in their
conditional abundance and less the patterns of occupancy. This suggests
that occupancy models alone may be inadequate for assessing risk from
anthropogenic disturbances and for describing the fine-scale
distribution of marine species (e.g., Winiarski \emph{et al.} 2014;
Flanders \emph{et al.} 2015). We suggest that while large-scale models
are useful to identify general geographic areas of import to sea ducks
(Silverman \emph{et al.} 2013; Flanders \emph{et al.} 2015), our
localized models provide more detailed estimates of sea duck
distribution and abundance within Nantucket Sound. In the context of
MSPs, large-scale models can help identify the areas to zone for OWED
while localized models may better inform the placement of the OWED
within the zoned area.

\emph{Environmental covariates that best explain sea duck distribution
and abundance.}---The biophysical associations with occupancy derived
from our models were relatively consistent among species, whereas their
associations with conditional abundance were more species-specific.
Distance to land, which often is positively associated with bathymetry,
often has a strong influence on sea duck occupancy estimates
(Guillemette \emph{et al.} 1993; Lewis \emph{et al.} 2008; Winiarski
\emph{et al.} 2014; Flanders \emph{et al.} 2015). Sediment grain size
can also have a strong influence on prey availability for foraging sea
ducks (Goudie \& Ankney 1988; Lovvorn \emph{et al.} 2009; Loring
\emph{et al.} 2013) and affected occupancy and conditional abundance in
this study. In addition, sea floor topographic variability also
influenced occupancy and conditional abundance, although the influence
of topography on prey availability is less understood. Sea surface
temperature and chlorophyll \emph{a} can have a strong influence on
occupancy estimates for sea ducks (Zipkin \emph{et al.} 2010; Flanders
\emph{et al.} 2015), although we found no effect of chlorophyll \emph{a}
during this study. Similarly, Zipkin \emph{et al.} (2010) found the
North Atlantic Oscillation (NAO) was important in all these species of
sea ducks at a continental scale, whereas we found no support that the
NAO affected the distribution of sea ducks at the scale of this study.
Certain covariates may associate with marine bird abundance or behavior
at specific scales and not at others (Logerwell \& Hargreaves 1996) and
this may explain the apparent discrepancy between studies in the effect
of chlorophyll \emph{a} and NAO.

The unexplained variation in our models and the predominance of marginal
spatiotemporal effects suggest that we probably omitted important
variable(s) relevant to the distribution of sea ducks in Nantucket
Sound. Moreover, we likely need better biophysical proxies for the
distribution of prey eaten by sea ducks or concurrent prey distribution
information (e.g., Vaitkus \& Bubinas 2001; Kaiser \emph{et al.} 2006;
Žydelis \emph{et al.} 2009; Cervencl \& Fernandez 2012; Cervencl
\emph{et al.} 2014), although this is typically considerably more
difficult to characterize at appropriate scales and does not guarantee
improved predictive accuracy (Grémillet \emph{et al.} 2008; Torres
\emph{et al.} 2008; Benoit-Bird \emph{et al.} 2013).

\emph{Marine Spatial Planning: where to place a wind farm in Nantucket
Sound given these estimates of the distribution and abundance of sea
ducks?}---In the past decade, ecosystem-based MSPs have become a reality
because comprehensive land use planning is a central component of
development plans in North America and Europe (Douvere 2008). One of the
biggest challenges facing marine spatial planners is the paucity of
relevant information on the spatial distribution and abundance of
biological resources including marine birds (Bradbury \emph{et al.}
2014; Flanders \emph{et al.} 2015). We developed fine-scale, spatially-
and temporally-explicit maps of the estimated distribution and abundance
of sea ducks in Nantucket Sound that could assist marine spatial
planners during the zoning process. At a continental scale, Nantucket
Sound regularly supports one of the largest concentrations of wintering
sea ducks in eastern North America (White \emph{et al.} 2009; Zipkin
\emph{et al.} 2010; Silverman \emph{et al.} 2013). Therefore if any
OWEDs are constructed in Nantucket Sound steps should be taken to
minimize impacts to this key wintering habitat. Bradbury \emph{et al.}
(2014) developed a sensitivity index that suggested that sea ducks were
particularly vulnerable to habitat displacement from foraging sites,
while they are less vulnerable to collision risk from OWEDs. Our models
suggest that the permitted OWED zone on Horseshoe Shoal is not located
in prime foraging habitat for most species of sea ducks, although large
numbers of eiders can use this area (see also Flanders \emph{et al.}
2015). Potential vulnerability to OWED has been incorporated into some
recent spatially-explicit planning efforts for seabirds (Garthe \&
Hüppop 2004; Winiarski \emph{et al.} 2014; Bradbury \emph{et al.} 2014)
though to our knowledge not yet for sea ducks.

\subsection{Acknowledgements}\label{acknowledgements}

We especially thank Simon Perkins, who had the primary role in
coordinating field logistics, and John Ambroult, who piloted the
majority of flights (and owned the plane). Special thanks to Andrea
Jones and Ellen Jedrey for their contributions to project design and
field work. Thanks also to Jack O'Brien and Phillip Kibler for piloting
flights as needed, and to Jasmine Smith-Gillen and numerous volunteer
observers and data recorders. Funding was provided by Foundation M, The
Island Foundation, Mass Audubon, Massachusetts Environmental Trust, the
State of Rhode Island, the U.S. Fish and Wildlife Service Sea Duck Joint
Venture, and the U.S. Fish and Wildlife Service Wildlife and Sport Fish
Restoration Program.

\subsection{References}\label{references}

\hangparas{16pt}{1}

\hypertarget{refs}{}
\hypertarget{ref-Austin2007}{}
Austin, M. (2007). Species distribution models and ecological theory: A
critical assessment and some possible new approaches. \emph{Ecological
Modelling}, \textbf{200}, 1--19.

\hypertarget{ref-Benoit-Bird2013}{}
Benoit-Bird, K.J., Battaile, B.C., Heppell, S.A., Hoover, B., Irons, D.,
Jones, N., Kuletz, K.J., Nordstrom, C.A., Paredes, R., Suryan, R.M.,
Waluk, C.M. \& Trites, A.W. (2013). Prey patch patterns predict habitat
use by top marine predators with diverse foraging strategies. \emph{PLoS
ONE}, \textbf{8}, e53348.

\hypertarget{ref-Borchers2002}{}
Borchers, D.L., Buckland, S.T. \& Zucchini, W. (2002). \emph{Estimating
animal abundance: Closed populations}. Springer, London.

\hypertarget{ref-Bowman2015}{}
Bowman, T.D., Silverman, E.D., Gilliland, S.G. \& Leirness, J.B. (2015).
Status and trends of North American sea ducks: Reinforcing the need for
better monitoring. \emph{Ecology and Conservation of North American Sea
Ducks. Studies in Avian Biology.} (eds J.-P.L. Savard, D.V. Derksen, D.
Esler \& J.M. Eadie), pp. 1--27. CRC Press, New York, NY, USA.

\hypertarget{ref-Bradbury2014}{}
Bradbury, G., Trinder, M., Furness, B., Banks, A.N., Caldow, R.W.G. \&
Hume, D. (2014). Mapping seabird sensitivity to offshore wind farms.
\emph{PLoS ONE}, \textbf{9}, e106366.

\hypertarget{ref-Breton2009}{}
Breton, S.-P. \& Moe, G. (2009). Status, plans and technologies for
offshore wind turbines in Europe and North America. \emph{Renewable
Energy}, \textbf{34}, 646--654.

\hypertarget{ref-Buhlmann2007}{}
Bühlmann, P. \& Hothorn, T. (2007). Boosting algorithms: Regularization,
prediction and model fitting. \emph{Statistical Science}, \textbf{22},
477--505.

\hypertarget{ref-Certain2008}{}
Certain, G. \& Bretagnolle, V. (2008). Monitoring seabirds population in
marine ecosystem: The use of strip-transect aerial surveys. \emph{Remote
Sensing of Environment}, \textbf{112}, 3314--3322.

\hypertarget{ref-Cervencl2012}{}
Cervencl, A. \& Fernandez, S.A. (2012). Winter distribution of Greater
Scaup \emph{Aythya marila} in relation to available food resources.
\emph{Journal of Sea Research}, \textbf{73}, 41--48.

\hypertarget{ref-Cervencl2014}{}
Cervencl, A., Troost, K., Dijkman, E., Jong, M. de, Smit, C.J., Leopold,
M.F. \& Ens, B.J. (2014). Distribution of wintering Common Eider
\emph{Somateria mollissima} in the Dutch Wadden Sea in relation to
available food stocks. \emph{Marine Biology}, \textbf{162}, 153--168.

\hypertarget{ref-Cunningham2005}{}
Cunningham, R.B. \& Lindenmayer, D.B. (2005). Modeling count data of
rare species: Some statistical issues. \emph{Ecology}, \textbf{86},
1135--1142.

\hypertarget{ref-Davis1997}{}
Davis, W.E., Jr. (1997). The Nantucket oldsquaw flight: New England's
greatest bird show? \emph{Bird Observer}, \textbf{25}, 16--22.

\hypertarget{ref-Douvere2008}{}
Douvere, F. (2008). The importance of marine spatial planning in
advancing ecosystem-based sea use management. \emph{Marine Policy},
\textbf{32}, 762--771.

\hypertarget{ref-Eilers1996}{}
Eilers, P.H.C. \& Marx, B.D. (1996). Flexible smoothing with B-splines
and penalties (with discussion). \emph{Statistical Science},
\textbf{11}, 89--121.

\hypertarget{ref-Fahrmeir2004}{}
Fahrmeir, L., Kneib, T. \& Lang, S. (2004). Penalized structured
additive regression for space-time data: A Bayesian perspective.
\emph{Statistica Sinica}, \textbf{14}, 731--761.

\hypertarget{ref-Flanders2015}{}
Flanders, N.P., Gardner, B., Winiarski, K.J., Paton, P.W.C., Allison, T.
\& O'Connell, A. (2015). Using a community occupancy model to identify
key seabird areas in southern New England. \emph{Marine Ecology Progress
Series}, \textbf{533}, 277--290.

\hypertarget{ref-Foley2010}{}
Foley, M.M., Halpern, B.S., Micheli, F., Armsby, M.H., Caldwell, M.R.,
Crain, C.M., Prahler, E., Rohr, N., Sivas, D., Beck, M.W., Carr, M.H.,
Crowder, L.B., Emmett Duffy, J., Hacker, S.D., McLeod, K.L., Palumbi,
S.R., Peterson, C.H., Regan, H.M., Ruckelshaus, M.H., Sandifer, P.A. \&
Steneck, R.S. (2010). Guiding ecological principles for marine spatial
planning. \emph{Marine Policy}, \textbf{34}, 955--966.

\hypertarget{ref-Friedman2001}{}
Friedman, J.H. (2001). Greedy function approximation: A gradient
boosting machine. \emph{The Annals of Statistics}, \textbf{29},
1189--1232.

\hypertarget{ref-Garthe2004}{}
Garthe, S. \& Hüppop, O. (2004). Scaling possible adverse effects of
marine wind farms on seabirds: Developing and applying a vulnerability
index. \emph{Journal of Applied Ecology}, \textbf{41}, 724--734.

\hypertarget{ref-Goudie1988}{}
Goudie, R.I. \& Ankney, C.D. (1988). Patterns of habitat use by sea
ducks wintering in southeastern Newfoundland. \emph{Ornis Scandinavica
(Scandinavian Journal of Ornithology)}, \textbf{19}, 249--256.

\hypertarget{ref-Gremillet2008}{}
Grémillet, D., Lewis, S., Drapeau, L., Van Der Lingen, C.D., Huggett,
J.A., Coetzee, J.C., Verheye, H.M., Daunt, F., Wanless, S. \& Ryan, P.G.
(2008). Spatial match--mismatch in the Benguela upwelling zone: Should
we expect chlorophyll and sea-surface temperature to predict marine
predator distributions? \emph{Journal of Applied Ecology}, \textbf{45},
610--621.

\hypertarget{ref-Guillemette1993}{}
Guillemette, M., Himmelman, J.H., Barette, C. \& Reed, A. (1993).
Habitat selection by common eiders in winter and its interaction with
flock size. \emph{Canadian Journal of Zoology}, \textbf{71}, 1259--1266.

\hypertarget{ref-Hastie1990}{}
Hastie, T.J. \& Tibshirani, R.J. (1990). \emph{Generalized additive
models}. CRC Press, Boca Raton, FL, USA.

\hypertarget{ref-Hoeting2009}{}
Hoeting, J.A. (2009). The importance of accounting for spatial and
temporal correlation in analyses of ecological data. \emph{Ecological
Applications}, \textbf{19}, 574--577.

\hypertarget{ref-HofHoth2015}{}
Hofner, B. \& Hothorn, T. (2015). Stabs: Stability selection with error
control, R package version 0.5-1.

\hypertarget{ref-Hofner2015}{}
Hofner, B., Boccuto, L. \& Göker, M. (2015a). Controlling false
discoveries in high-dimensional situations: Boosting with stability
selection. \emph{BMC Bioinformatics}, \textbf{16}, 144.

\hypertarget{ref-Hofner2011}{}
Hofner, B., Hothorn, T., Kneib, T. \& Schmid, M. (2011). A framework for
unbiased model selection based on boosting. \emph{Journal of
Computational and Graphical Statistics}, \textbf{20}, 956--971.

\hypertarget{ref-Hofner2015b}{}
Hofner, B., Mayr, A. \& Schmid, M. (2015b). gamboostLSS: An R package
for model building and variable selection in the GAMLSS framework.
\emph{Accepted for publication in Jorunal of Statistical Software}.

\hypertarget{ref-Hofner2015a}{}
Hofner, B., Mayr, A., Fenske, N. \& Schmid, M. (2015c). gamboostLSS:
Boosting methods for GAMLSS models, R package version 1.1-3.

\hypertarget{ref-Hofner2014}{}
Hofner, B., Mayr, A., Robinzonov, N. \& Schmid, M. (2014). Model-based
boosting in R: A hands-on tutorial using the R package mboost.
\emph{Computational Statistics}, \textbf{29}, 3--35.

\hypertarget{ref-Hothorn2015}{}
Hothorn, T., Bühlmann, P., Kneib, T., Schmid, M. \& Hofner, B. (2015).
Mboost: Model-based boosting, R package version 2.4-2.

\hypertarget{ref-Hothorn2010}{}
Hothorn, T., Bühlmann, P., Kneib, T., Schmid, M. \& Hofner, B. (2010).
Model-based boosting 2.0. \emph{Journal of Machine Learning Research},
\textbf{11}, 2109--2113.

\hypertarget{ref-Johnson1980}{}
Johnson, D.H. (1980). The comparison of usage and availability
measurements for evaluating resource preference. \emph{Ecology},
\textbf{61}, 65--71.

\hypertarget{ref-Johnson2006}{}
Johnson, C.J., Nielsen, S.E., Merrill, E.H., Mcdonald, T.L. \& Boyce,
M.S. (2006). Resource selection functions based on use--availability
data: Theoretical motivation and evaluation methods. \emph{Journal of
Wildlife Management}, \textbf{70}, 347--357.

\hypertarget{ref-Johnson2004}{}
Johnson, C.J., Seip, D.R. \& Boyce, M.S. (2004). A quantitative approach
to conservation planning: Using resource selection functions to map the
distribution of mountain caribou at multiple spatial scales.
\emph{Journal of Applied Ecology}, \textbf{41}, 238--251.

\hypertarget{ref-Kaiser2006}{}
Kaiser, M.J., Galanidi, M., Showler, D.A., Elliott, A.J., Caldow,
R.W.G., Rees, E.I.S., Stillman, R.A. \& Sutherland, W.J. (2006).
Distribution and behaviour of Common Scoter \emph{Melanitta nigra}
relative to prey resources and environmental parameters. \emph{Ibis},
\textbf{148}, 110--128.

\hypertarget{ref-Kinlan2012}{}
Kinlan, B., Menza, C. \& Huettmann, F. (2012). Chapter 6: Predictive
modeling of seabird distribution patterns in the New York Bight. \emph{A
biogeographic assessment of seabirds, deep sea corals and ocean habitats
of the New York Bight: Science to support offshore spatial planning.
NOAA Technical Memorandum NOS NCCOS 141} (eds C. Menza, B. Kinlan, D.
Dorfman, M. Poti \& C. Caldow), pp. 87--224. NOAA, Silver Spring, MD.

\hypertarget{ref-Kneib2009}{}
Kneib, T., Hothorn, T. \& Tutz, G. (2009). Variable selection and model
choice in geoadditive regression models. \emph{Biometrics}, \textbf{65},
626--634.

\hypertarget{ref-Kneib2008}{}
Kneib, T., Müller, J. \& Hothorn, T. (2008). Spatial smoothing
techniques for the assessment of habitat suitability.
\emph{Environmental and Ecological Statistics}, \textbf{15}, 343--364.

\hypertarget{ref-Langston2013}{}
Langston, R.H.W. (2013). Birds and wind projects across the pond: A UK
perspective. \emph{Wildlife Society Bulletin}, \textbf{37}, 5--18.

\hypertarget{ref-Larsen2007}{}
Larsen, J.K. \& Guillemette, M. (2007). Effects of wind turbines on
flight behaviour of wintering common eiders: Implications for habitat
use and collision risk. \emph{Journal of Applied Ecology}, \textbf{44},
516--522.

\hypertarget{ref-Lewis2008}{}
Lewis, T.L., Esler, D. \& Boyd, W.S. (2008). Foraging Behavior of Surf
Scoters (\emph{Melanitta perspicillata}) and White-Winged Scoters
(\emph{M. fusca}) in relation to clam density: Inferring food
availability and habitat quality. \emph{The Auk}, \textbf{125},
149--157.

\hypertarget{ref-Logerwell1996}{}
Logerwell, E.A. \& Hargreaves, N.B. (1996). The distribution of sea
birds relative to their fish prey off Vancouver Island: Opposing results
at large and small spatial scales. \emph{Fisheries Oceanography},
\textbf{5}, 163--175.

\hypertarget{ref-Loring2013}{}
Loring, P.H., Paton, P.W.C., McWilliams, S.R., McKinney, R.A. \& Oviatt,
C.A. (2013). Densities of wintering scoters in relation to benthic prey
assemblages in a North Atlantic estuary. \emph{Waterbirds}, \textbf{36},
144--155.

\hypertarget{ref-Lovvorn2009}{}
Lovvorn, J.R., Grebmeier, J.M., Cooper, L.W., Bump, J.K. \& Richman,
S.E. (2009). Modeling marine protected areas for threatened eiders in a
climatically changing Bering Sea. \emph{Ecological Applications},
\textbf{19}, 1596--1613.

\hypertarget{ref-Maloney2012}{}
Maloney, K.O., Schmid, M. \& Weller, D.E. (2012). Applying additive
modelling and gradient boosting to assess the effects of watershed and
reach characteristics on riverine assemblages. \emph{Methods in Ecology
and Evolution}, \textbf{3}, 116--128.

\hypertarget{ref-Martin2005}{}
Martin, T.G., Wintle, B.A., Rhodes, J.R., Kuhnert, P.M., Field, S.A.,
Low-Choy, S.J., Tyre, A.J. \& Possingham, H.P. (2005). Zero tolerance
ecology: Improving ecological inference by modelling the source of zero
observations. \emph{Ecology Letters}, \textbf{8}, 1235--1246.

\hypertarget{ref-Mayr2012}{}
Mayr, A., Fenske, N., Hofner, B., Kneib, T. \& Schmid, M. (2012).
Generalized additive models for location, scale and shape for
high-dimensional data - a flexible approach based on boosting.
\emph{Journal of the Royal Statistical Society: Series C (Applied
Statistics)}, \textbf{61}, 403--427.

\hypertarget{ref-Meinshausen2010}{}
Meinshausen, N. \& Bühlmann, P. (2010). Stability selection (with
discussion). \emph{Journal of the Royal Statistical Society: Series B
(Statistical Methodology)}, \textbf{72}, 417--473.

\hypertarget{ref-Musial2010}{}
Musial, W. \& Ram, B. (2010). \emph{Large-scale offshore wind power in
the United States: Assessment of opportunities and barriers}. National
Renewable Energy Laboratory, Golden, CO, USA.

\hypertarget{ref-Nagelkerke1991}{}
Nagelkerke, N.J.D. (1991). A note on a general definition of the
coefficient of determination. \emph{Biometrika}, \textbf{78}, 691--692.

\hypertarget{ref-Nur2011}{}
Nur, N., Jahncke, J., Herzog, M.P., Howar, J., Hyrenbach, K.D., Zamon,
J.E., Ainley, D.G., Wiens, J.A., Morgan, K., Ballance, L.T. \&
Stralberg, D. (2011). Where the wild things are: Predicting hotspots of
seabird aggregations in the California Current System. \emph{Ecological
Applications}, \textbf{21}, 2241--2257.

\hypertarget{ref-Polasky2008}{}
Polasky, S., Nelson, E., Camm, J., Csuti, B., Fackler, P., Lonsdorf, E.,
Montgomery, C., White, D., Arthur, J., Garber-Yonts, B., Haight, R.,
Kagan, J., Starfield, A. \& Tobalske, C. (2008). Where to put things?
Spatial land management to sustain biodiversity and economic returns.
\emph{Biological Conservation}, \textbf{141}, 1505--1524.

\hypertarget{ref-Punt2009}{}
Punt, M.J., Groeneveld, R.A., van Ierland, E.C. \& Stel, J.H. (2009).
Spatial planning of offshore wind farms: A windfall to marine
environmental protection? \emph{Ecological Economics}, \textbf{69},
93--103.

\hypertarget{ref-RCoreTeam2014}{}
R Core Team. (2014). \emph{R: A language and environment for statistical
computing.} R Foundation for Statistical Computing, Vienna, Austria.

\hypertarget{ref-Richards2008}{}
Richards, S.A. (2008). Dealing with overdispersed count data in applied
ecology. \emph{Journal of Applied Ecology}, \textbf{45}, 218--227.

\hypertarget{ref-Rigby2005}{}
Rigby, R.A. \& Stasinopoulos, D.M. (2005). Generalized additive models
for location, scale and shape (with discussion). \emph{Journal of the
Royal Statistical Society: Series C (Applied Statistics)}, \textbf{54},
507--554.

\hypertarget{ref-Santora2004}{}
Santora, C., Hade, N. \& Odell, J. (2004). Managing offshore wind
developments in the United States: Legal, environmental and social
considerations using a case study in Nantucket Sound. \emph{Ocean \&
Coastal Management}, \textbf{47}, 141--164.

\hypertarget{ref-Schmid2008}{}
Schmid, M. \& Hothorn, T. (2008). Boosting additive models using
component-wise P-splines. \emph{Computational Statistics \& Data
Analysis}, \textbf{53}, 298--311.

\hypertarget{ref-Shah2013}{}
Shah, R.D. \& Samworth, R.J. (2013). Variable selection with error
control: Another look at stability selection. \emph{Journal of the Royal
Statistical Society: Series B (Statistical Methodology)}, \textbf{75},
55--80.

\hypertarget{ref-Silverman2013}{}
Silverman, E.D., Saalfeld, D.T., Leirness, J.B. \& Koneff, M.D. (2013).
Wintering sea duck distribution along the Atlantic Coast of the United
States. \emph{Journal of Fish and Wildlife Management}, \textbf{4},
178--198.

\hypertarget{ref-Torres2008}{}
Torres, L.G., Read, A.J. \& Halpin, P. (2008). Fine-scale habitat
modeling of a top marine predator: Do prey data improve predictive
capacity. \emph{Ecological Applications}, \textbf{18}, 1702--1717.

\hypertarget{ref-Vaitkus2001}{}
Vaitkus, G. \& Bubinas, A. (2001). Modelling of sea duck spatial
distribution in relation to food resources in Lithuanian offshore waters
under the gradient of winter climatic conditions. \emph{Acta Zoologica
Lituanica}, \textbf{11}, 288--302.

\hypertarget{ref-White2009}{}
White, T.P., Veit, R.R. \& Perry, M.C. (2009). Feeding ecology of
Long-Tailed Ducks \emph{Clangula hyemalis} wintering on the Nantucket
Shoals. \emph{Waterbirds}, \textbf{32}, 293--299.

\hypertarget{ref-Winiarski2014}{}
Winiarski, K.J., Miller, D.L., Paton, P.W. \& McWilliams, S.R. (2014). A
spatial conservation prioritization approach for protecting marine birds
given proposed offshore wind energy development. \emph{Biological
Conservation}, \textbf{169}, 79--88.

\hypertarget{ref-Wood2006}{}
Wood, S.N. (2006). \emph{Generalized additive models: An introduction
with R}. Chapman and Hall/CRC, Boca Raton, FL, USA.

\hypertarget{ref-Zeileis2008}{}
Zeileis, A., Kleiber, C. \& Jackman, S. (2008). Regression models for
count data in R. \emph{Journal of Statistical Software}, \textbf{27},
1--25.

\hypertarget{ref-Zipkin2010}{}
Zipkin, E.F., Gardner, B., Gilbert, A.T., Jr, A.F.O., Royle, J.A. \&
Silverman, E.D. (2010). Distribution patterns of wintering sea ducks in
relation to the North Atlantic Oscillation and local environmental
characteristics. \emph{Oecologia}, \textbf{163}, 893--902.

\hypertarget{ref-Zydelis2009}{}
Žydelis, R., Esler, D., Kirk, M. \& Sean Boyd, W. (2009). Effects of
off-bottom shellfish aquaculture on winter habitat use by molluscivorous
sea ducks. \emph{Aquatic Conservation: Marine and Freshwater
Ecosystems}, \textbf{19}, 34--42.

\newpage
\subsection{Figure Legends}\label{figure-legends}

\textbf{Figure 1.} Actual aerial strip transect tracks (gray lines)
during winter (October - April, 2003 - 2005) sea duck surveys (n = 30)
in Nantucket Sound, Massachusetts, US. The grid indicates the extent of
the 1100 km\textsuperscript{2} study area and its division into 504
2.25km\textsuperscript{2} segments. The polygon in northwest Nantucket
Sound indicates the 62 km\textsuperscript{2} area of permitted wind
energy development on Horseshoe Shoal.

\textbf{Figure 2.} Marginal functional plots for stably selected
covariates in the occupancy (probability of presence) and conditional
abundance (mean and overdispersion of count models) of Common Eider
(COEI), scoter (SCOT), and Long-tailed Duck (LTDU) in Nantucket Sound
during three winters, 2003 - 2005. Plots illustrate the partial
contribution to the additive predictor (Y-axis) of a covariate holding
all other covariates at their mean. Within a model, univariate plots
(i.e., lines) share a Y-axis scale, enabling direct comparisons of
effect sizes among covariates and species. For bivariate plots, the
Y-axis and X-axis reflect the first and second variables listed in the
interaction, respectively; colors indicate the direction and magnitude
of the partial contribution (blacks = negative, reds = positive; darker
colors = larger effect) and are likewise comparable within a model.
Northing by easting effects are given only at 31 December. For factor
variables, only the general association (i.e., positive or negative)
with the additive predictor is given. Covariate abbreviations correspond
to Equation 1.

\textbf{Figure 3.} Occupancy probability for Common Eider (COEI), scoter
(SCOT), and Long-tailed Duck (LTDU) in Nantucket Sound during three
winters, 2003 - 2005. Occupancy probabilities (top row) represent the
median expected probability of sea duck presence in a 1.5 km x ca. 180 m
transect through a given segment predicted on 10 evenly-spaced dates
from 15 November through 1 April in each winter. Spatiotemporal
variation in occupancy (\%; bottom row) is indicated by the median
absolute deviation, MAD, of occupancy probability relative to the
median. Predicted values are categorized based on their quartiles;
segments with the highest occupancy or variability (values \(\geq\) 98th
percentile) are outlined in black.

\textbf{Figure 4.} Conditional abundance of Common Eider (COEI), scoter
(SCOT), and Long-tailed Duck (LTDU) in Nantucket Sound during three
winters, 2003 - 2005. Conditional abundances (top row) represent the
median expected number of sea ducks, assuming their presence, in a 1.5
km x ca. 180 m transect in each segment predicted on 10 evenly-spaced
dates from 15 November through 1 April in each winter. Spatiotemporal
variation in conditional abundance (\%; bottom row) is indicated by the
median absolute deviation, MAD, relative to the median. Predicted values
are categorized based on their quartiles; segments with the highest
conditional abundance or variability (values \(\geq\) 98th percentile)
are outlined in black.

\textbf{Figure 5.} Unconditional abundance of Common Eider (COEI),
scoter (SCOT), and Long-tailed Duck (LTDU) in Nantucket Sound during
three winters, 2003 - 2005. Median abundances (top row) represent the
expected number of sea ducks along a 1.5 km x ca. 180 m transect within
each segment predicted on 10 evenly-spaced dates from 15 November
through 1 April in each winter. Spatiotemporal variation in abundance
(\%; bottom row) is estimated from the median absolute deviation, MAD,
relative to the median. Predicted values are categorized based on their
quartiles; segments with the highest abundance or variability (values
\(\geq\) 98th percentile) are outlined in black.

\textbf{Figure 6.} Relationship between observed and predicted total
abundance of Common Eider (COEI), scoter (SCOT), and Long-tailed Duck
(LTDU) during 30 aerial surveys of Nantucket Sound over three winters,
2003 - 2005. The dashed line indicates a 1:1 relationship between
predicted and observed abundances in surveyed segments; points below and
above this line indicate underestimates and overestimates of predicted
abundances, respectively.

\newpage

\includegraphics{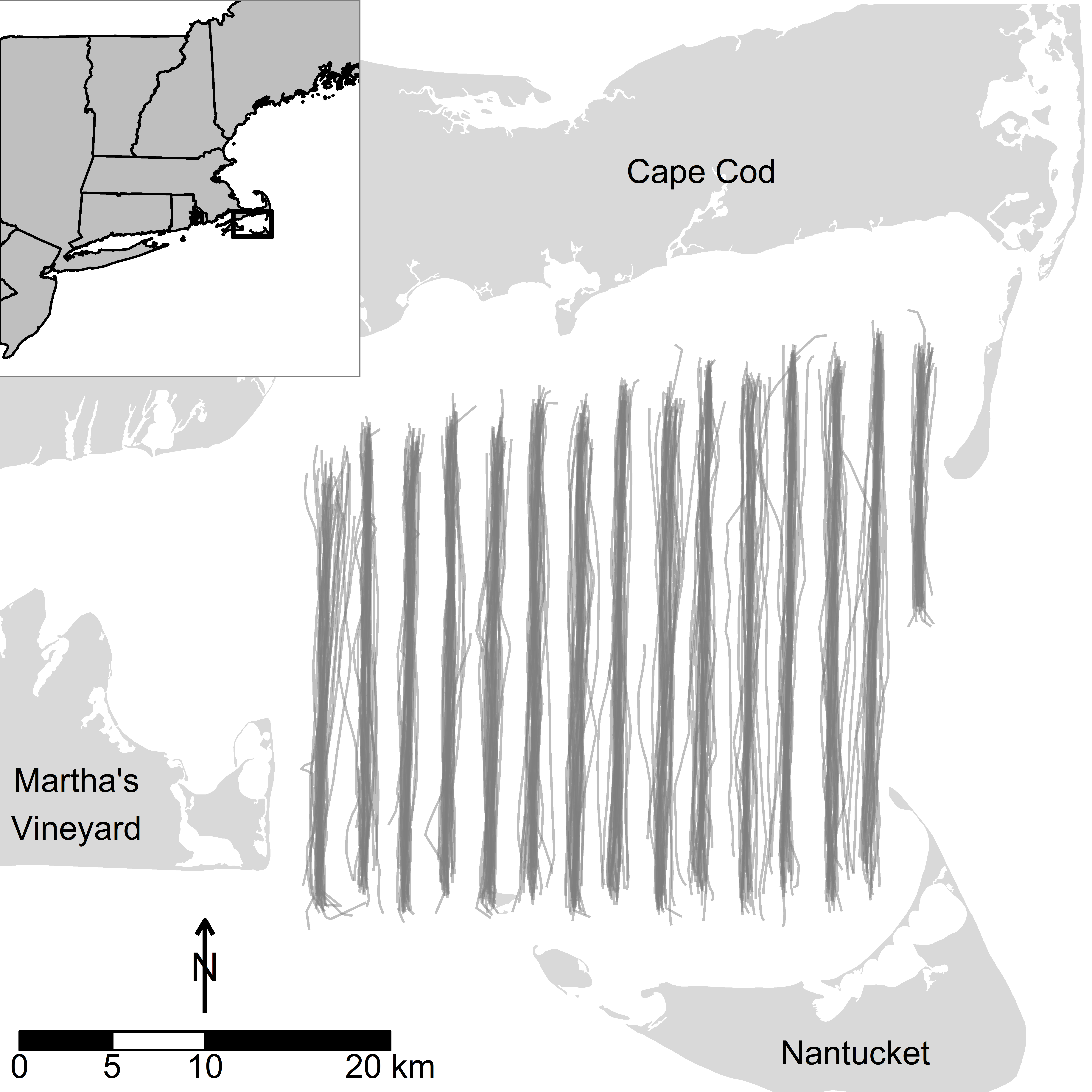}\\
\textbf{Figure 1}

\newpage

\includegraphics{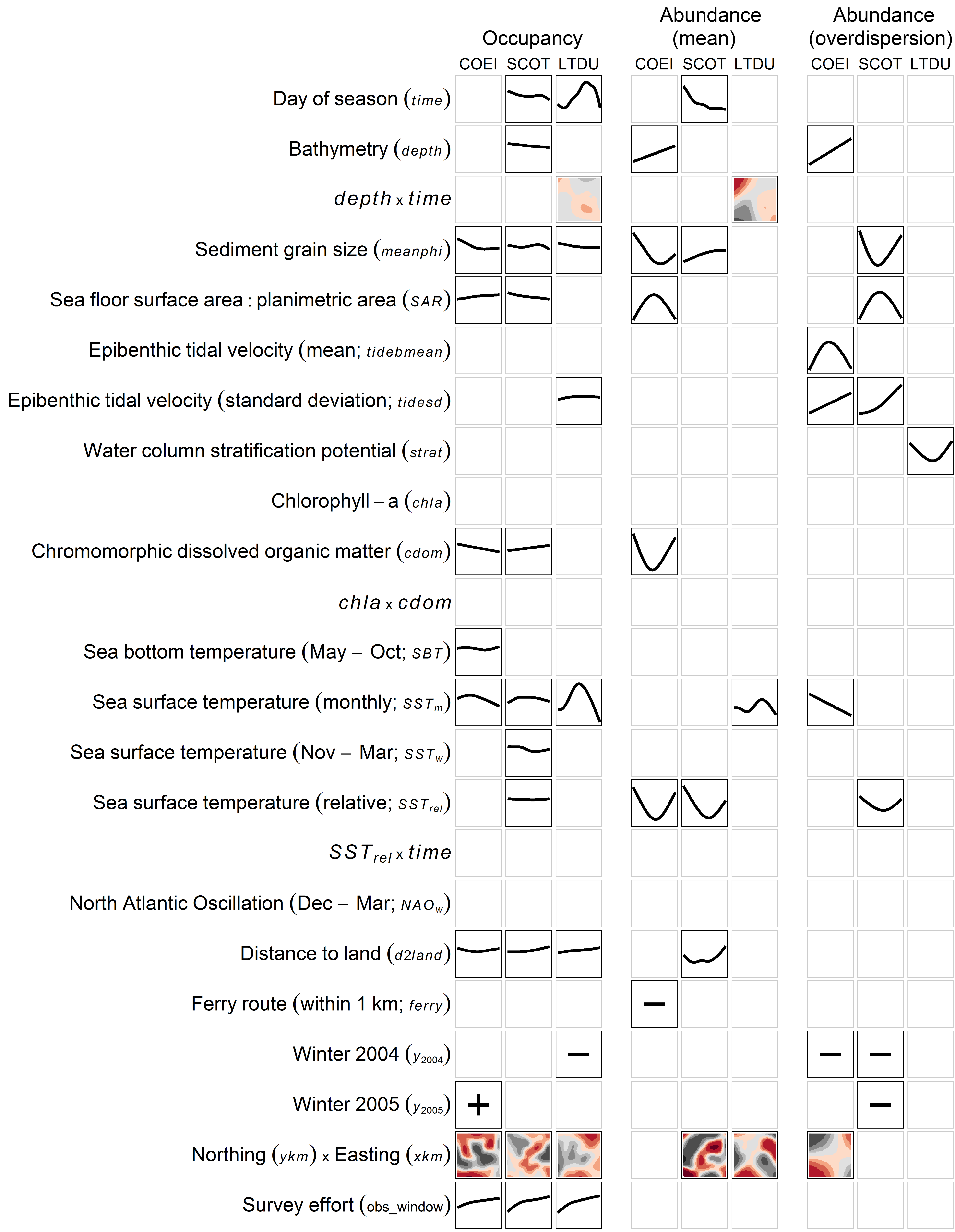}\\
\textbf{Figure 2}

\newpage

\includegraphics{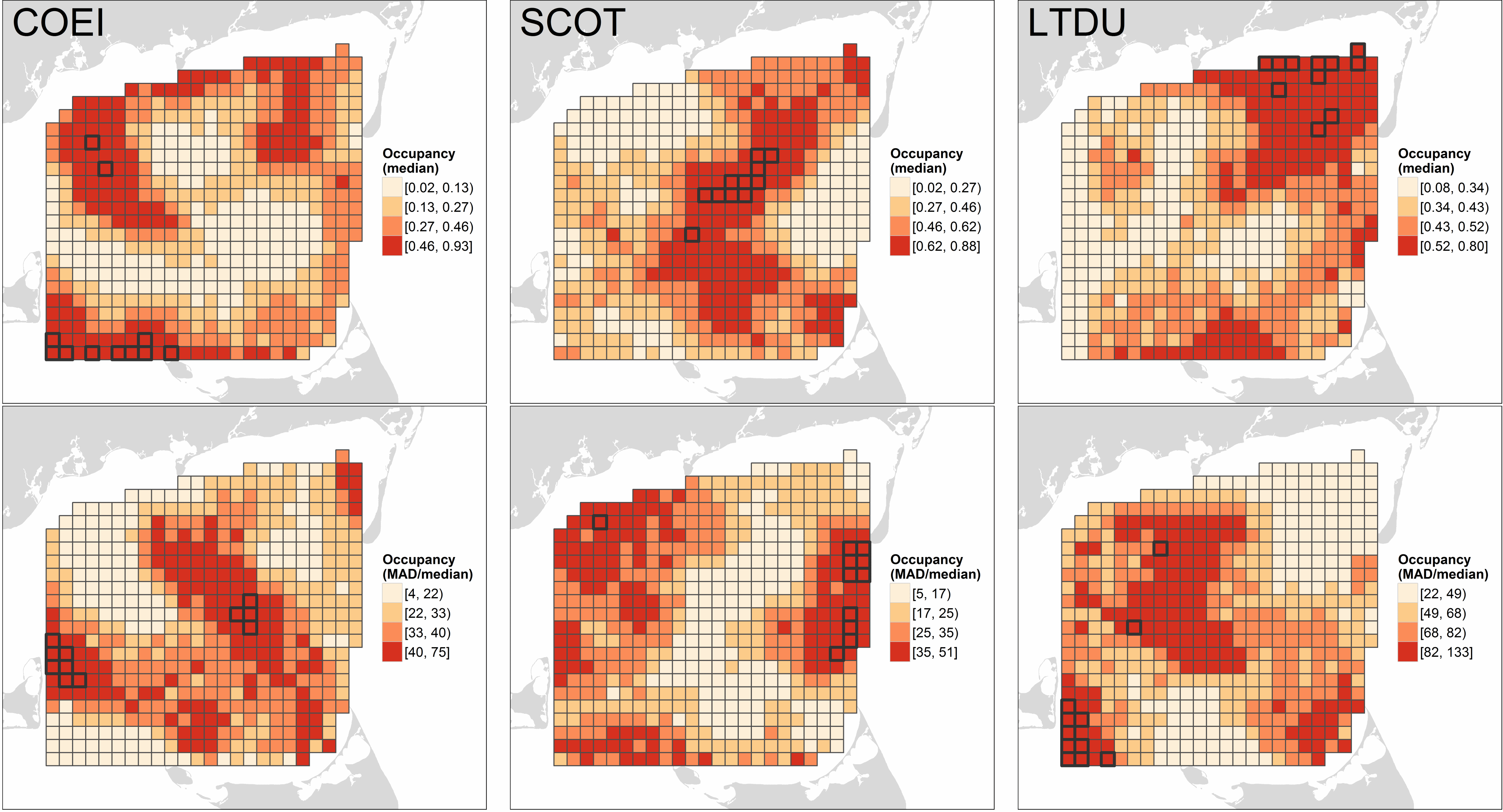}\\
\textbf{Figure 3}

\newpage

\includegraphics{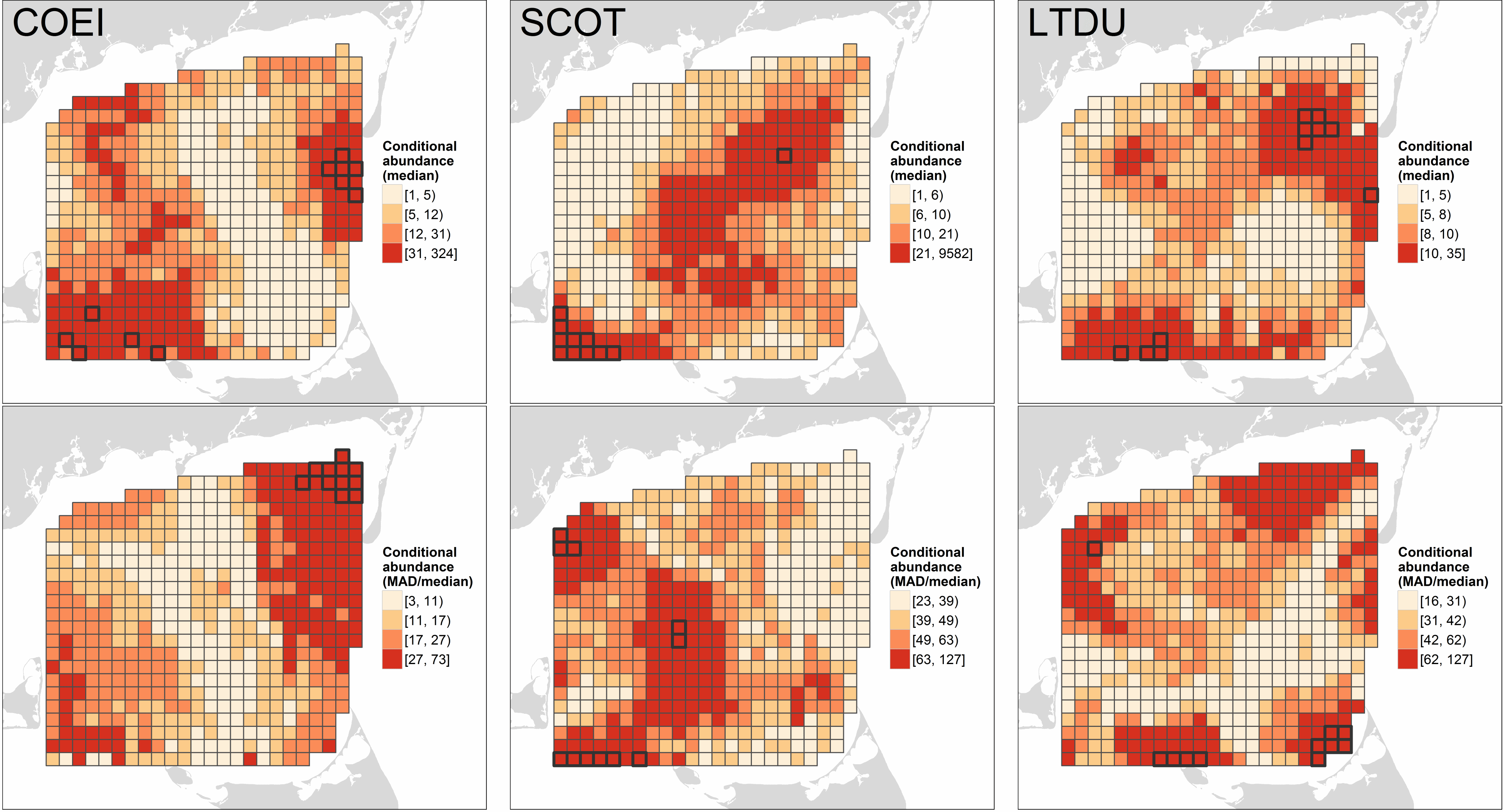}\\
\textbf{Figure 4}

\newpage

\includegraphics{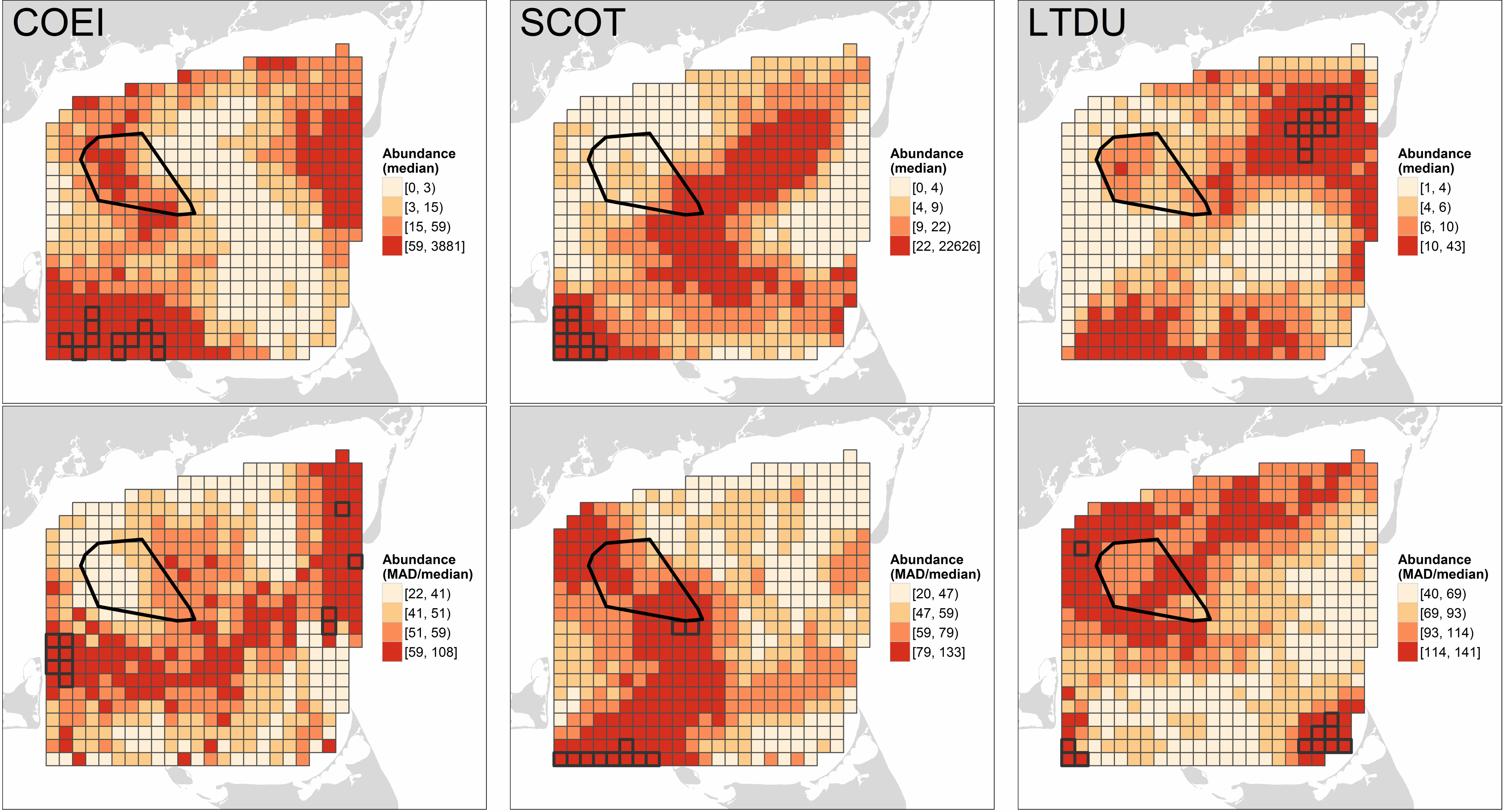}\\
\textbf{Figure 5}

\newpage

\includegraphics{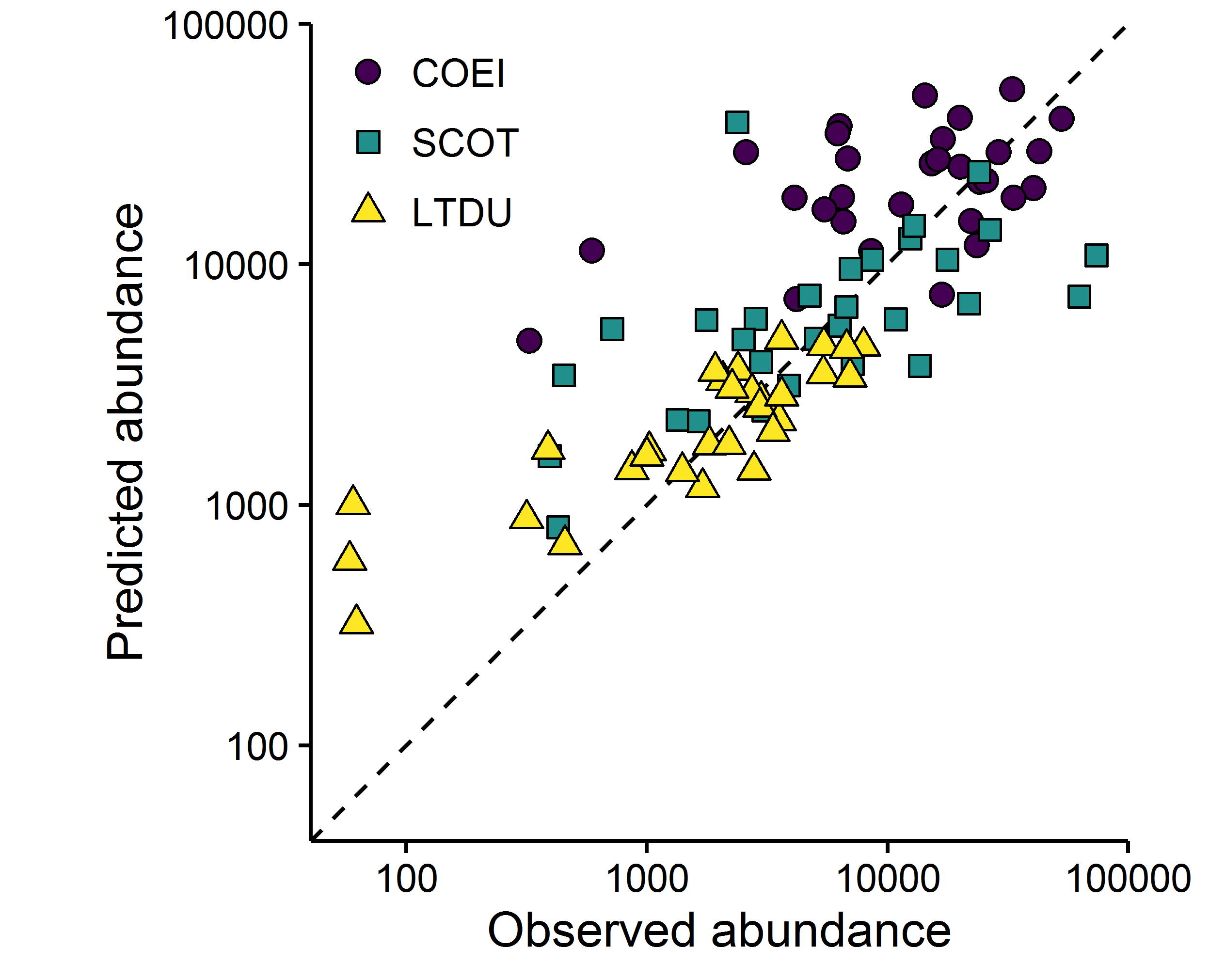}\\
\textbf{Figure 6}

\end{document}